\documentclass[twocolumn,
pra
,superscriptaddress
,floatfix
,aps
,10pt
,showpacs]{revtex4-1}

\usepackage{graphicx}
\usepackage{amssymb}
\usepackage{amsmath}
\usepackage{color}
\newcommand{\ve}[1]{{\mathbf #1}}
\newcommand{\vect}[1]{{\mathbf #1}}
\newcommand{\Frac}[2]{\displaystyle\frac{#1}{#2}}

\begin{document}

%\linenumbers

%\title{Dipolar fermions in a multi-layer geometry}
\title{Dipolar fermions in a multilayer geometry}

\author{M. Callegari}
%\email{mig.callegari@gmail.com} 
\affiliation{Department of Physics and Astronomy ``Galileo Galilei'',
  University of Padova, Via Marzolo 8, 35131 Padova, Italy }
\affiliation{Departamento de F\'isica Te\'orica de la Materia
  Condensada \& Condensed Matter Physics Center (IFIMAC), Universidad
  Aut\'onoma de Madrid, Madrid 28049, Spain}

\author{M. M. Parish}
\email{meera.parish@monash.edu}
\affiliation{London Centre for Nanotechnology, Gordon Street, London,
  WC1H 0AH, United Kingdom}
\affiliation{School of Physics \& Astronomy, Monash University,
  Victoria 3800, Australia}

\author{F. M. Marchetti}
\email{francesca.marchetti@uam.es} 
\affiliation{Departamento de F\'isica Te\'orica de la Materia
  Condensada \& Condensed Matter Physics Center (IFIMAC), Universidad
  Aut\'onoma de Madrid, Madrid 28049, Spain}

%\date{\today}
\date{January 26, 2016}

\begin{abstract}
  We investigate the behavior of identical dipolar fermions with
  aligned dipole moments in two-dimensional multilayers at zero
  temperature. We consider density instabilities that are driven by
  the attractive part of the dipolar interaction and, for the case of
  bilayers, we elucidate the properties of the stripe phase recently
  predicted to exist in this interaction regime. When the number of
  layers is increased, we find that this ``attractive'' stripe phase
  exists for an increasingly larger range of dipole angles, and if the
  interlayer distance is sufficiently small, the stripe phase
  eventually spans the full range of angles, including the situation
  where the dipole moments are aligned perpendicular to the planes.
  In the limit of an infinite number of layers, we derive an analytic
  expression for the interlayer effects in the density-density
  response function and, using this result, we find that the stripe
  phase is replaced by a collapse of the dipolar system.
\end{abstract}


% 67.85.Lm 	Degenerate Fermi gases
% 71.45.Gm 	Exchange, correlation, dielectric and 
%               magnetic response functions, plasmons
% 71.45.Lr 	Charge-density-wave systems 
%               (see also 75.30.Fv Spin-density waves)
% 68.65.Ac 	Multilayers (in Low-dimensional, mesoscopic, 
%               nanoscale and other related systems: 
%               structure and nonelectronic properties)

\maketitle

%%%%%%%%%%%%%%%%%%%%%%%%%%%%%%%%%%%%%%%%%%%%%%%%%%%
\section{Introduction}
Motivated by the prospect of novel many-body phases generated by
anisotropic long-range dipolar interactions, much attention has
recently been devoted to ultracold polar molecules and magnetic
atoms~\cite{Baranov2008,lahaye2009,Baranov_review2012}. Dipolar
fermions, in particular, can be used to simulate strongly correlated
phenomena in electron systems, including charge density modulations
(stripes) and unconventional
superconductivity~\cite{CDW_review,kivelson1998,kivelson2003}.

Quantum degeneracy has already been achieved for dipolar Fermi gases
of atoms with a permanent magnetic dipole moment such as
chromium~\cite{Naylor_PRA2015}, dysprosium~\cite{Lu_2012} and
erbium~\cite{Aikawa_2014b}. This has enabled the observation of
dipole-driven Fermi surface deformations in the Fermi liquid
phase~\cite{Aikawa_2014}.
However, to investigate many-body phenomena at stronger dipole-dipole
interactions, it appears necessary to use polar molecules, which
generally possess larger dipole moments --- the electric dipole moment
can be as large as 5.5~Debye in the case of
$^{133}$Cs$^{6}$Li~\cite{Carr2009}. Thus far, there has been major
progress towards producing quantum degenerate clouds of long-lived
fermionic dipolar molecules using
$^{40}$K$^{87}$Rb~\cite{Ni2008,Ospelkaus2010,Ni2010},
$^{23}$Na$^{6}$Li~\cite{Heo_2012},
$^{133}$Cs$^{6}$Li~\cite{Repp_2013}, and
$^{23}$Na$^{40}$K~\cite{Wu_2012,Park_2015}.

\begin{figure}
\centering
\includegraphics[width=0.8\linewidth,angle=0]{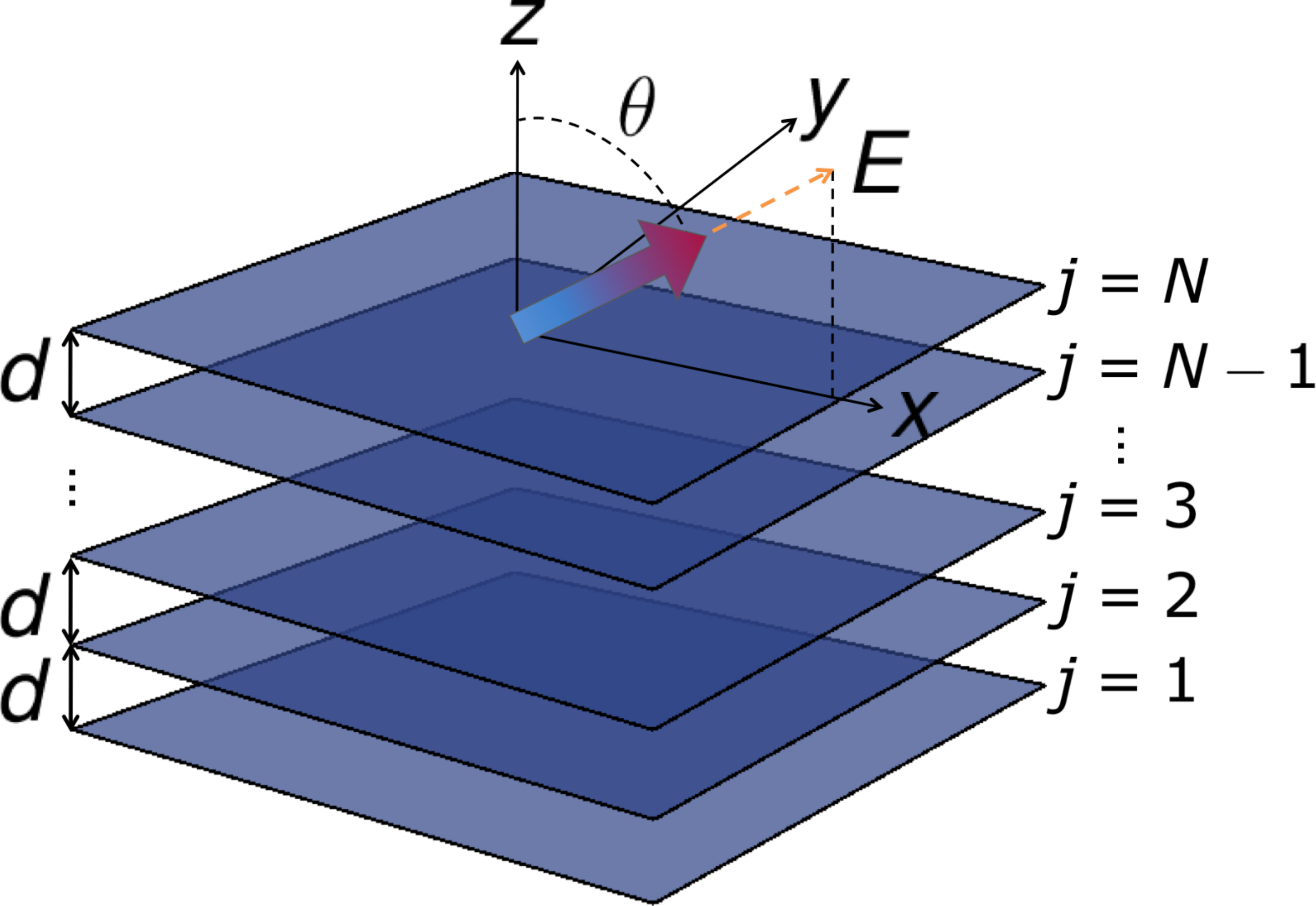}
\caption{(Color online) Schematic representation of the system
  geometry. A gas of dipolar fermions is confined to $N$ 2D layers
  labelled by an index $j=1, 2, \dots, N$ and separated by a distance
  $d$. We assume that each layer has the same density of dipoles  $n$.
  The dipole moment of each fermion is aligned by an electric field
  $\vect{E} = (E_x,0,E_z)$ in the $x$-$z$ plane ($\varphi=0$
  direction), at an angle $\theta$ with respect to the
  $\hat{\vect{z}}$ direction.}
\label{fig:setup}
\end{figure}

The dipole-dipole interaction can be further tuned and enhanced by
confining the polar molecules to \textit{two-dimensional} (2D) layers.
Such a geometry has been used to suppress chemical
reactions~\cite{miranda2011} and to stabilize the gas against
mechanical collapse, which arises in three dimensions for a
sufficiently strong
interaction~\cite{Miyakawa_PRA_2008,Sogo2009,Endo_PRA_2010,Zhang_PRA_2009,Zhang_PRA_2010,Krieg_PRA_2015,Baillie_PRA_2015}.
Furthermore, by aligning all the dipole moments with a strong electric
field, the nature of the effective 2D dipolar interaction within the
plane may be externally manipulated: The dipole-dipole repulsion is
maximized by aligning the dipole moments perpendicular to the plane,
while anisotropy and attraction are gradually introduced by varying
the dipole tilt (see Fig.~\ref{fig:setup}).
This possibility has stimulated much theoretical work on dipolar
fermionic gases in single and multilayer 2D
geometries~\cite{Baranov_review2012}.

For the single-layer geometry and for small (but non-zero) tilting
angles, the weakly interacting system corresponds to a Landau Fermi
liquid with deformed Fermi surface~\cite{Lu2012}, similarly to the
case in 3D.
With increasing dipolar interaction (or cloud density), the system is
then predicted to undergo a transition to a uni-directional density
modulated
phase~\cite{yamaguchi2010,DasSarma2010,babadi2011,sieberer2011}, where
the modulations are perpendicular to the direction of the dipole tilt.
Such a ``stripe'' phase has also been shown to exist in the
\emph{isotropic} case where the dipoles are aligned perpendicular to
the layer, thus requiring the system to spontaneously break the
rotational symmetry~\cite{parish2012}.
This result has recently been supported by density functional theory
calculations~\cite{vanZyl_2015}, which predict a transition to a
stripe phase followed by a transition to a triangular Wigner crystal
at higher coupling. Quantum Monte Carlo calculations also find a
Wigner crystal phase for the case of perpendicularly aligned dipoles,
although at a much higher dipolar interaction than that obtained from
density functional theory~\cite{Matveeva_2012}.

For tilting angles greater than a critical angle, the attractive part
of the dipolar interaction can lead to $p$-wave superfluidity in the
single-layer system~\cite{bruun2008}, and this phase may even coexist
with stripe order~\cite{Wu_2015}.
A sufficiently strong attraction eventually drives a mechanical
instability of the cloud towards
collapse~\cite{bruun2008,yamaguchi2010,sieberer2011,parish2012,block_PRB_2014}.
However, interestingly, if one instead considers a bilayer geometry,
the additional layer stabilizes the collapse at large tilt angles ---
as long as the dipoles are aligned out of the plane ($\theta < \pi/2$)
--- to form a new stripe phase, where the density modulations are
oriented along the direction of the dipole tilt~\cite{marchetti_13}.

In this article, we investigate such a stripe phase, which is
generated by the attractive part of the dipolar interaction for large
enough dipole tilt angle. We start by elucidating its properties in
the case of the bilayer, where we provide a classical argument for how
the stripes in each layer are shifted with respect to each other. Then
we extend our results for the density-density response function to the
multilayer geometry. We employ an approach based on the STLS
scheme~\cite{STLSpaper,parish2012,marchetti_13} that incorporates
exchange interactions only, which should be reasonable for the
``attractive'' stripe phase~\cite{marchetti_13}, although note that we
neglect the possibility of pairing~\cite{pikovski2010,Matveeva_2014}
and other stripe phases driven by the strong
repulsion~\cite{block2012,marchetti_13}.
As the number of layers $N$ is increased, we find that the attractive
stripe phase spans an increasingly larger region of the phase diagram.
However, this stripe phase eventually gives way to collapse in the $N
\to \infty$ limit.

The paper is organized as follows: In Sec.~\ref{sec:model}, we
describe the system geometry and introduce the STLS scheme which
allows us to evaluate the density-density response function matrix in
the multilayer geometry; in Sec.~\ref{sec:phize}, we describe the
properties of the density instabilities driven by the attractive part
of the interlayer dipolar interaction and, in Sec.~\ref{sec:class}, we
explain via a classical model how the stripes in each layer are
shifted with respect to each other. In Sec.~\ref{sec:finit} we extend
the results to a generic number of layers $N$, while, in
Sec.~\ref{sec:infin} we consider the $N \to \infty$ limit. The
concluding remarks are gathered in Sec.~\ref{sec:concl}.

%%%%%%%%%%%%%%%%%%%%%%%%%%%%%%%%%%%%%%%%%%%%%%%%%%%
\section{Multilayer system and model}
\label{sec:model}
We consider a gas of polar fermionic molecules in a multilayer
geometry, as shown in the schematic picture in
Fig.~\ref{fig:setup}. The molecules have a dipole moment $D$ and are
confined to $N$ two-dimensional layers, each labelled by an index $j =
1, 2, \dots, N$, and equally separated by a distance $d$.
We assume that the dipoles are aligned by an external electric field
$\vect{E}=(E_x,0,E_z)$ in the $x$-$z$ plane, which is tilted at an
angle $\theta$ with respect to the $\hat{\vect{z}}$ direction. Within
each layer, we parameterize the $x$-$y$ in-plane wavevector by polar
coordinates $\ve{q} = (q,\varphi)$, where $\varphi = 0$ corresponds to
the direction of the dipole tilt.

In the limit $qW \ll 1$, where $W$ is the layer width, the effective
2D intralayer interaction between dipoles takes the following
form~\cite{fischer_06}:
\begin{equation}
  v_{jj} (\ve{q}) \equiv v (\ve{q}) = V_0 - 2\pi D^2 q
  \xi(\theta,\varphi)\; ,
\label{eq:intra}
\end{equation}
where $\xi(\theta,\varphi) = \cos^2\theta - \sin^2\theta
\cos^2\varphi$. Here, $\ve{q}$ corresponds to the relative wavevector
between two dipoles.
The constant $V_0$ is a short-range contact interaction term that in
general depends on the width $W$~\cite{fischer_06}, %which acts as
yielding a natural UV cut-off.  Since we are considering identical
fermions, the system properties will not depend on $V_0$.

In the limit where the layer width is much smaller than the layer
separation, $W \ll d$, the interaction between two dipoles in
different layers $j>l$ is given by~\cite{Li_Hwang_DasSarma10}:
\begin{equation} 
  v_{jl}(\ve{q}) = -2\pi D^2 q e^{-(j-l)q d} \left[\xi(\theta,\varphi)
    + i \sin 2\theta \cos \varphi \right] \; .
\label{eq:inter}
\end{equation}
The remaining interlayer interactions can be obtained from the
condition $v_{lj}(\ve{q}) = v_{jl}(-\ve{q}) = v^*_{jl}(\ve{q})$, which
is derived from the fact the the dipolar interaction is always real in
real space.  Likewise, the momentum-space interaction is complex for
$\theta \neq 0$ since the real-space interaction is not invariant
under the transformation $\ve{r} \mapsto - \ve{r}$.

Assuming that each layer has the same density $n$, we define the Fermi
wavevector $k_F = \sqrt{4\pi n}$. This allows us to define the
dimensionless interaction strength $U = mD^2k_F/\hbar^2$, where $m$ is
the fermion mass. The other parameters that can be independently
varied are the dipole tilt angle $\theta$ and the dimensionless layer
separation $k_F d$.

%%%%%%%%%%%%%%%%%%%%%%%%%%%%%%%%%%%%%%%%%%%%%%%%%%%
\subsection{Response function and STLS equations}
\label{sec:stlse}
Similarly to Ref.~\cite{marchetti_13}, we make use of linear response
theory to analyze density wave instabilities. In the multilayer
system, the linear density response $\delta n$ to an external
perturbing potential $V^{ext}$ defines the density-density response
function matrix~\cite{macdonald_94},
\begin{align}
  \delta n_j(\ve{q},\omega) = \sum_l \chi_{jl}(\ve{q},\omega)
  V_l^{ext}(\ve{q},\omega) \; ,
\end{align}
where $j, l$ are the layer indices. A divergence in the static
density-density response function matrix $\chi_{jl}(\ve{q},\omega=0)$
signals an instability of the system.  Specifically, the system is
unstable towards forming a stripe phase when the smallest eigenvalue
$\tilde{\chi}_{\text{min}}(\ve{q})$ of the static response function
matrix first diverges at a critical value of the wavevector
$\vect{q}_c = (q_c,\varphi_c)$.

While the response function is known exactly for the non-interacting
gas, typically one can only incorporate the effect of interactions
approximately.  A standard approach is the random phase approximation
(RPA), where one replaces the external potential with one that
contains an effective potential due to the perturbed density:
$V_j^{ext} \mapsto V_j^{ext} + \sum_l v_{jl} \delta n_l$. However, RPA
neglects exchange correlations, which are always important in the
dipolar system, even in the long-wavelength limit $q\to
0$~\cite{parish2012}.
This issue may be remedied using a conserving Hartree-Fock
approximation~\cite{babadi2011,sieberer2011,block2012}, but we choose
a simpler and physically motivated approach where correlations are
included via local field factors $G_{jl}(\ve{q})$ \cite{vignale_book}.
This yields the inverse density-density response function matrix
\begin{equation} 
  \hat{\chi}^{-1} {}_{jl} (\ve{q},\omega) =
  \Frac{\delta_{jl}}{\Pi(\ve{q},\omega)} - v_{jl}(\ve{q})
  \left[1-G_{jl} (\ve{q}) \right] \; ,
\label{eq:dende}
\end{equation}
where $\Pi (\ve{q},\omega)$ is the non-interacting response function,
which, for equal density layers, reads as~\cite{stern67,macdonald_94}
\begin{equation*}
  \Pi(\ve{q},i\omega) = \frac{m}{2\pi \hbar^2 b} \left\{\sqrt{2} \left[a +
      \sqrt{a^2 + \left(\omega b\right)^2} \right]^{1/2} -b\right\}\; ,
\end{equation*}
with $a=\frac{b^2}{4} - b\frac{k_F^2}{m} - \omega^2$ and
$b=\frac{q^2}{m}$. RPA corresponds to taking the limit where the
layer-resolved local field factors $G_{jl} (\ve{q})$ in
Eq.~\eqref{eq:dende} are all zero.

The response function~\eqref{eq:dende} can be related to the
layer-resolved static structure factor $S_{jl} (\ve{q})$ by the
fluctuation-dissipation theorem:
\begin{equation}
  S_{jl} (\ve{q}) = -\frac{\hbar}{\pi n} \int_0^{\infty} d\omega
  \chi_{jl}(\ve{q},i\omega)\; .
\label{eq:struc}
\end{equation}
In the non-interacting limit, the static structure factor is diagonal,
i.e., $S_{jl}^{(0)} (\vect{q}) = \delta_{jl} \ S_{}^{(0)} (q)$, and
can be evaluated exactly (see App.~\ref{sec:nissf}), where
\begin{equation}
  S_{}^{(0)} (q) = \frac{2}{\pi} \arcsin \left(\Frac{q}{2 k_F}\right)
  + \frac{q}{\pi k_F} \sqrt{1 - \left(\frac{q}{2 k_F}\right)^2}\; ,
\label{eq:struz}
\end{equation}
for $q \leq 2k_F$, while $S_{}^{(0)} (q) = 1$ for $q >
2k_F$~\cite{macdonald_94,vignale_book}.

To determine the local field factors, we consider the STLS
approximation scheme, where we have the
expression~\cite{STLSpaper,macdonald_94}:
\begin{equation}
  G_{jl} (\ve{q}) = \frac{1}{n} \int \frac{d\ve{k}}{(2\pi)^2}
  \frac{\ve{q}\cdot\ve{k}}{q^2} \frac{v_{jl} (\ve{k})}{v_{jj}
    (\ve{q})} \left[\delta_{jl} - S_{jl} (\ve{q}-\ve{k}) \right] \; .
\label{eq:local}
\end{equation}
In principle, $G_{jl} (\ve{q})$ can be determined self-consistently by
solving Eqs.~\eqref{eq:dende}, \eqref{eq:struc} and
\eqref{eq:local}. Note that this self-consistent approach %in principle
includes all correlations beyond RPA and is not just limited to
exchange correlations. As such, the STLS scheme has proven to be a
powerful method for treating strongly correlated electron systems such
as the 2D electron gas~\cite{vignale_book}.
We previously adapted an improved version of this scheme to the
dipolar system, both in the single-~\cite{parish2012} and
double-layer~\cite{marchetti_13} geometries. The scheme is improved by
imposing, at each iteration step, the condition that the intralayer
pair correlation function is zero at zero distance, $g_{jj} (0)=0$,
where,
\begin{equation}
  g_{jl}(\ve{r}) = 1+ \frac{1}{n} \int \frac{d\ve{q}}{(2\pi)^2}
  e^{i\ve{q}\cdot\ve{r}} \left[S_{jl}(\ve{q}) - \delta_{jl} \right] \;
  .
\label{eq:pairc}
\end{equation}
This ensures that the intralayer static structure factor $S_{jj}
(\ve{q})$ is dominated by Pauli exclusion in the long wavelength
limit, $q \gg 2k_F$, and that the system response is independent of
the short-range contact interaction term $V_0$ and the cut-off $W$.

In the following, we first review the bilayer case and describe the
instability to a stripe phase occurring for large tilt angles
$\theta$, where the modulations are oriented along the dipole tilt,
i.e., along $\varphi=0$. We then show how the instability to the
$\varphi = 0$ stripe phase can be well described using exchange
correlations only, and we use this to investigate its existence in the
multilayer geometry.

%%%%%%%%%%%%%%%%%%%%%%%%%%%%%%%%%%%%%%%%%%%%%%%%%%%
\section{The $\varphi=0$ stripe phase in bilayers}
\label{sec:phize}
The case of two layers ($N=2$) was previously analysed within the STLS
self-consistent approximation scheme in
Ref.~\cite{marchetti_13}. Here, at sufficiently small tilt angles
$\theta<\theta_c$, and by increasing the value of the dimensionless
coupling strength $U$, there is an instability from the uniform phase
to a stripe phase with modulations along the $y$-axis ($\varphi=\pi/2$
stripe phase).
The instability to this stripe phase is driven by intralayer
correlations beyond exchange, which are induced by the repulsive part of
the intralayer interaction potential $v(\vect{q})$.
By contrast, for $\theta_c <\theta<\pi/2$, the system develops an
instability to a stripe phase along the $x$-axis ($\varphi=0$ stripe
phase). While for a single layer, the attractive sliver of the
intralayer interaction produces a collapse of the dipolar Fermi gas at
large tilt angles, the bilayer geometry stabilizes the collapse in
favour of a $\varphi=0$ stripe phase, which thus derives from a
competition between the intralayer attraction in the $\varphi=0$
direction and the interlayer interaction.

Interestingly, this latter stripe phase can be accurately described
using intralayer exchange correlations only. In fact, it was found for
this phase that the intralayer pair correlation function $g_{jj}
(\vect{r})$ deviated only slightly from the non-interacting case,
while the interlayer correlation function $g_{12} (\vect{r}) \sim
1$. In terms of local correlations, the interlayer local field factor
can thus be neglected, $G_{12} (\vect{r}) = 0$, while the intralayer
one $G_{jj} (\vect{q}) \equiv G(\vect{q})$ is determined from the
non-interacting intralayer structure factor~\eqref{eq:struz}:
\begin{equation}
  G (\ve{q}) = \frac{1}{n} \int \frac{d\ve{k}}{(2\pi)^2}
  \frac{\ve{q}\cdot\ve{k}}{q^2} \frac{v (\ve{k})}{v (\ve{q})} \left[1
    - S_{}^{(0)} (q) \right] \; .
\label{eq:locaz}
\end{equation}
We refer to this approximation scheme as the exchange-only STLS
approximation (X-STLS).
For the single-layer case~\cite{parish2012}, this approach yielded an
instability towards collapse at large $\theta$ that agreed with the
predictions from Hartree-Fock
calculations~\cite{bruun2008,yamaguchi2010,babadi2011,sieberer2011}.

\begin{figure}
\centering
\includegraphics[width=1.0\linewidth,angle=0]{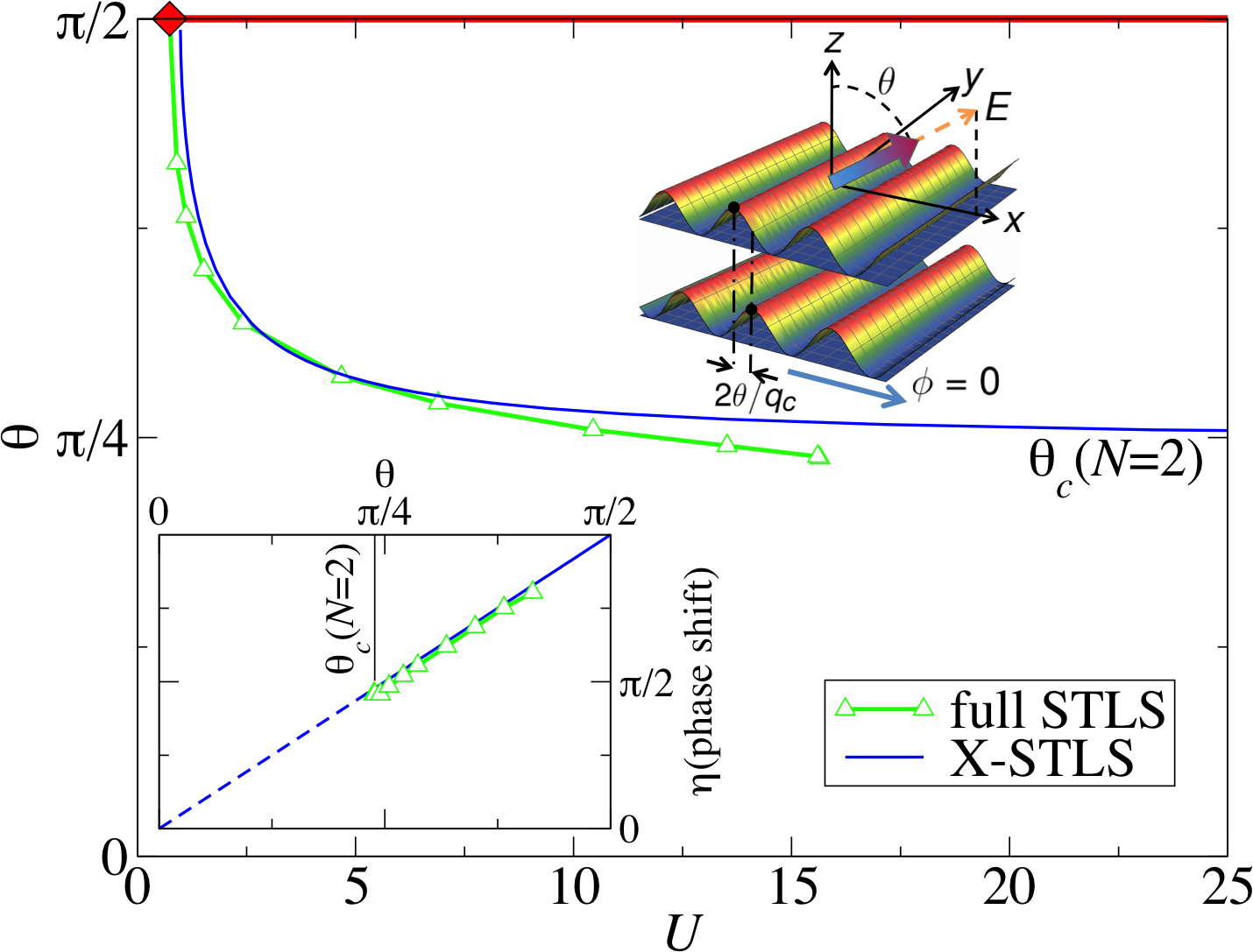}
\caption{(Color online) Main panel: Phase boundary for the $\varphi=0$
  stripe phase in the two-layer geometry as a function of the tilt
  angle $\theta$ and the dimensionless interaction strength $U$ at
  fixed interlayer distance $k_F d = 2$. Density modulations are in
  the direction $\varphi=0$ of the dipole tilt (schematic figure). The
  instability to this phase is to the right of the plotted boundaries:
  The full STLS results~\cite{marchetti_13} ([green] open triangles)
  are compared to the results obtained with exchange-only correlations
  ([blue] solid line), also referred to as the X-STLS approximation
  scheme. Within X-STLS, the $\varphi=0$ stripe phase appears for
  $\theta>\theta_c \simeq 0.79$. At $\theta=\pi/2$, the gas is
  unstable towards mechanical collapse for $U \gtrsim 1.57$ ([red]
  diamond symbol and thick solid line), where the gas compressibility
  is infinite. The density modulations in the two layers have a phase
  shift $\eta$ (lower inset) equal to $2 \theta$, i.e., the shift
  between the modulations $2\theta/q_c$ (schematic figure).}
\label{fig:phase}
\end{figure}

The phase boundary between the normal phase and the $\varphi=0$ stripe
phase obtained from the full STLS scheme is displayed in
Fig.~\ref{fig:phase} ([green] open triangles) and is compared with the
results of the X-STLS approximation ([blue] solid line). The value of
the critical tilt angle for such a phase is $\theta_c \simeq 0.75$ if
evaluated within the full STLS scheme, while it is slightly higher,
$\theta_c \simeq 0.79$, if evaluated within the X-STLS
approximation. It turns out that, for the phase boundary, the X-STLS
approximation works particularly well at large angles, all the way up
to $\theta=\pi/2$, where, for $U \gtrsim 1.57$ ([red] diamond symbol
and thick solid line), the gas collapses because the Fermi pressure is
not high enough to counteract the strong dipolar attraction.

The eigenvectors of the density-density response
function~\eqref{eq:dende} determine the phase-shift $\eta$ between
layers, and for the bilayer geometry we have found
that~\cite{marchetti_13}:
\begin{equation}
  e^{i \eta} = -\Frac{v_{12} (\vect{q}) [1 - G_{12}
      (\vect{q})]}{|v_{12} (\vect{q}) [1 - G_{12} (\vect{q})]|} \; .
\end{equation}
For both $\varphi=\pi/2$ and $\varphi=0$ stripe phases, we find that
the interlayer phase shift between the modulations is independent of
the dipole interaction strength $U$ and the layer distance $d$.
In particular, for the $\varphi=\pi/2$ stripe phase, both the
interaction and the local field factor are real and thus both layer
modulations are always in phase, i.e., $\eta=0$. On the other hand,
for the $\varphi=0$ stripe phase, if we consider an exchange-only
approximation (X-STLS) for which $G_{12}(\vect{q}) = 0$, we obtain a
phase shift of $\eta = 2\theta$.
The phase shift for the $\varphi=0$ stripe phase is plotted in the
inset of Fig.~\ref{fig:phase}. We see a very good agreement between
the full STLS results ([green] open triangles) and the simplified
X-STLS scheme ([blue] solid line).
The $\varphi=0$ result may at first appear counter-intuitive, but it
can be reproduced by evaluating the classical interaction energy
between an infinite layer of dipoles in one layer and a single dipole
in the second layer, as we discuss next.

%%%%%%%%%%%%%%%%%%%%%%%%%%%%%%%%%%%%%%%%%%%%%%%%%%%
\subsection{Classical model}
\label{sec:class}
\begin{figure}
\centering
\includegraphics[width=0.4\linewidth,angle=0]{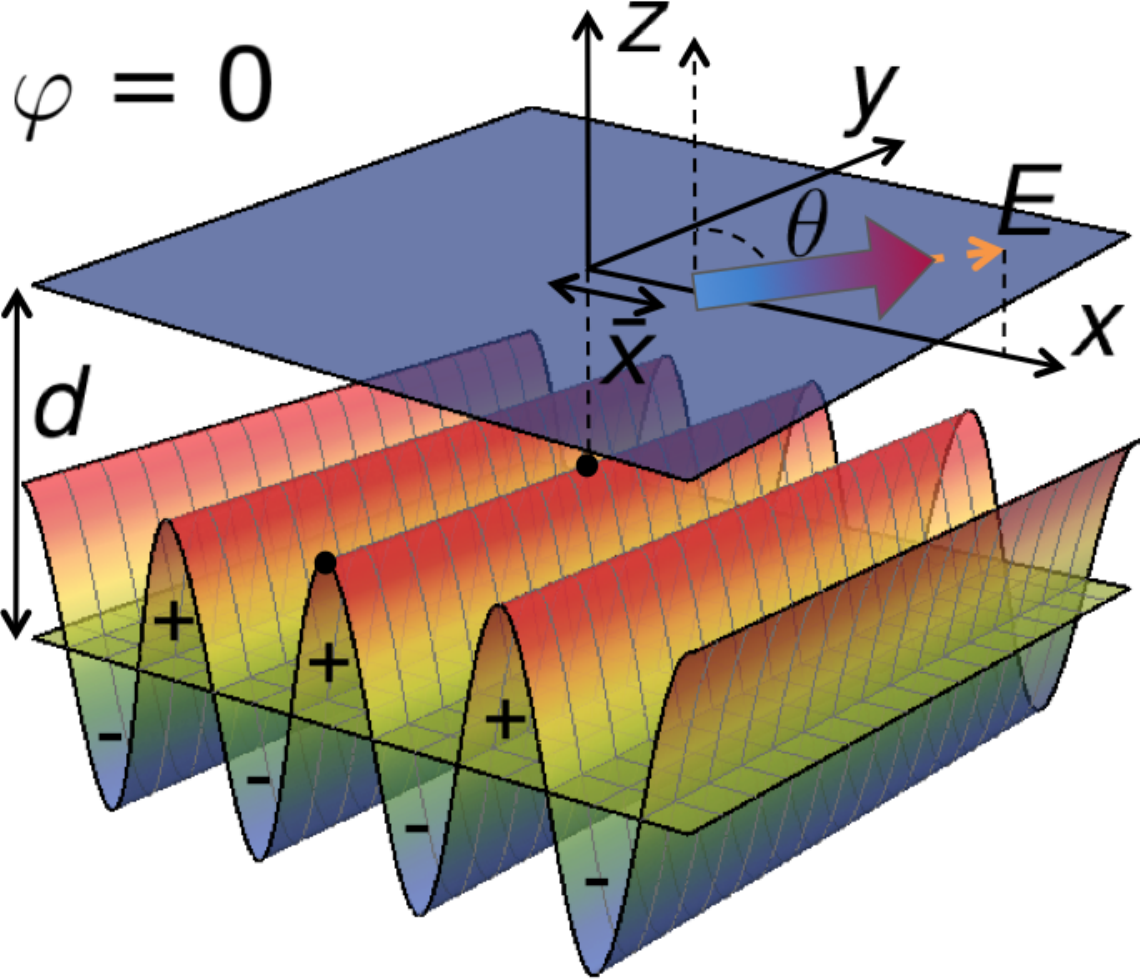} \hspace{25pt}
\includegraphics[width=0.4\linewidth,angle=0]{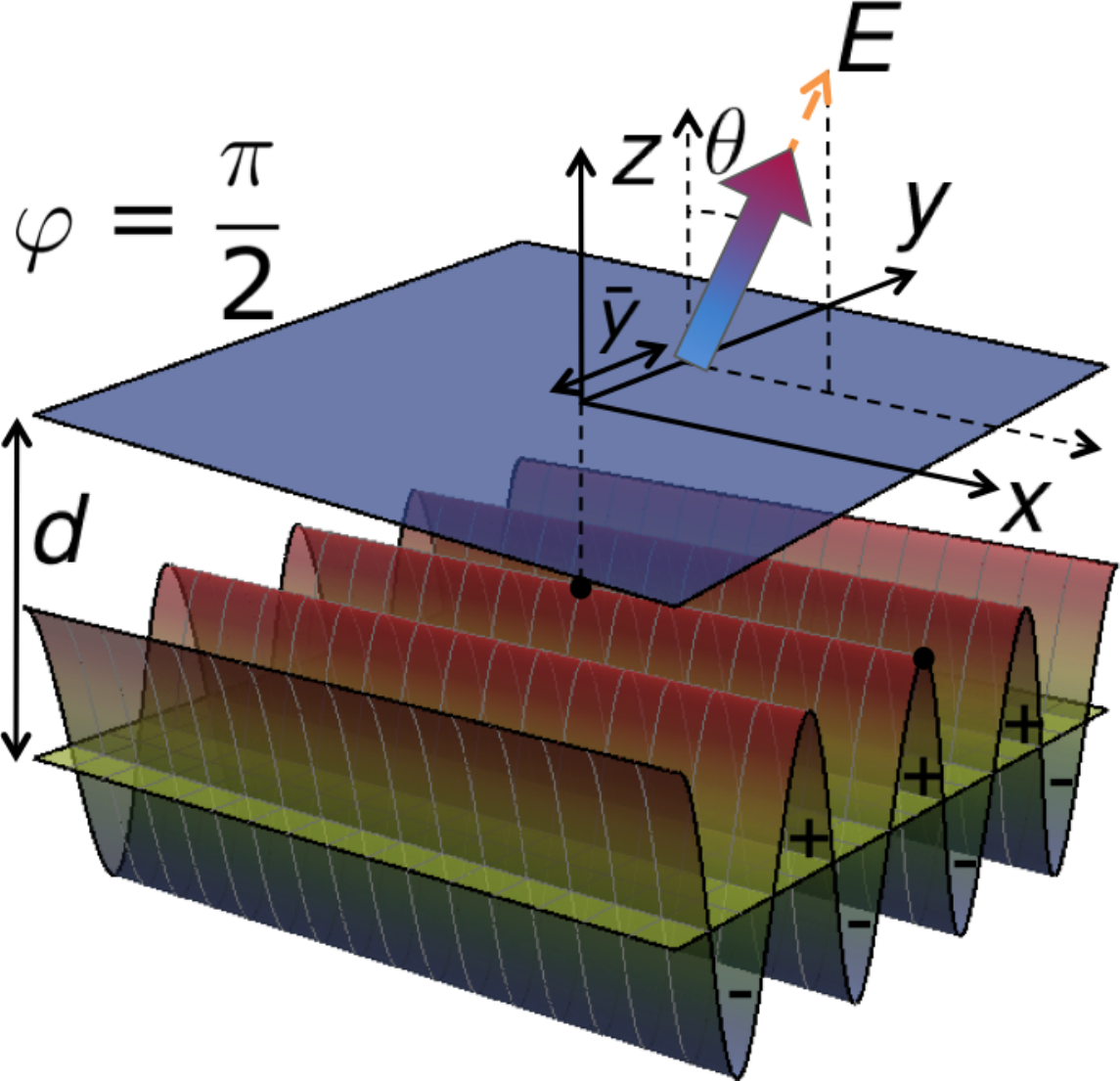}
\caption{(Color online) Schematic representation of a single dipole in
  the top layer interacting classically with an infinite bottom layer
  of dipoles arranged in a stripe modulated phase.  As in
  Fig.~\ref{fig:setup}, all dipoles are aligned by an electric field
  $\mathbf{E}$. In the left (right) panel, the single dipole is
  shifted by $\bar{x}$ ($\bar{y}$) with respect to one of the stripe
  crests for the $\varphi=0$ ($\varphi=\frac{\pi}{2}$) stripe phase.}
\label{fig:class}
\end{figure}
For the bilayer geometry, a simple classical model can easily explain
the phase shifts found in both $\varphi=0$ and $\varphi=\pi/2$ stripe
phases. Let us consider the simplified case of an infinite layer of
dipoles whose density is modulated sinusoidally with a wavevector $q_c
= 2\pi/\lambda_c$ and amplitude $\rho_0$. We further assume that this
interacts classically with a single dipole positioned at $\vect{r}_0$
in the other layer. The two layer densities are thus respectively
given by:
\begin{align}
  \rho^{(1)} (\vect{r}) &= n + \rho_0 \cos (\vect{q}_c \cdot \vect{r})
  \\
  \rho^{(2)} (\vect{r}')&= \delta(\vect{r}' - \vect{r}_0) \; .
\end{align}
For the $\varphi=0$ ($\varphi=\pi/2$) stripe phase we have
$\hat{\vect{q}}_c = \hat{\vect{x}}$ ($\hat{\vect{q}}_c =
\hat{\vect{y}}$) and $\vect{r}_0 = (\bar{x},0)$ ($\vect{r}_0 =
(0,\bar{y})$) --- see the schematic representation of both geometries
in Fig.~\ref{fig:class}.
The classical interaction energy is given by
\begin{equation} 
  E_{cl} = \int d\vect{r} d\vect{r}' \rho^{(1)} (\vect{r}) v_{12}
  (\vect{r} - \vect{r}') \rho^{(2)} (\vect{r}')\; ,
\label{eq:class}
\end{equation}
where 
\begin{equation*}
  v_{12} (\vect{r}) = D^2 \frac{x^2 + y^2 + d^2 - 3 (x\sin\theta +
    d\cos\theta)^2}{(x^2 + y^2 + d^2)^{5/2}}
\end{equation*}
is the interlayer potential (assuming that the distance $d$ is much
larger than the layer thickness $W$). For a uniform density
distribution $\rho^{(1)} (\vect{r}) = n$, the classical interaction
energy would be zero; therefore only deviations from the average
density in $\rho^{(1)} (\vect{r})$ contribute (either positively or
negatively) to $E_{cl}$.

Considering the Fourier transforms $\rho^{(1,2)} (\vect{r}) = \int
\frac{d\ve{q}}{(2 \pi)^2} \rho^{(1,2)} (\vect{q}) e^{i \ve{q} \cdot
  \ve{r}}$ and $v_{12} (\vect{r}) = \int \frac{d\ve{q}}{(2 \pi)^2}
v_{12} (\vect{q}) e^{i \ve{q} \cdot \ve{r}}$, we can rewrite
\eqref{eq:class} to obtain
\begin{equation}
  E_{cl} = \int \Frac{d\vect{q}}{(2 \pi)^2} \rho^{(1)} (-\vect{q})
  v_{12} (\vect{q}) \rho^{(2)} (\vect{q})\; ,
\end{equation}
where
\begin{align*}
  \rho^{(1)} (\vect{q}) &= (2\pi)^2 \left[n \delta(\ve{q}) + \rho_0
    \Frac{\delta(\vect{q} + \vect{q}_c) +\delta(\vect{q} -
      \vect{q}_c)}{2} \right]\\
  \rho^{(2)} (\vect{q})&= e^{- i \vect{q} \cdot \vect{r}_0} \; .
\end{align*}
Thus, as $v_{12} (0) = 0$, we get in general
\begin{equation}
  E_{cl} = \Frac{\rho_0}{2} \left[v_{12} (-\vect{q}_c) e^{i \vect{q}_c
      \cdot \vect{r}_0} + v_{12} (\vect{q}_c) e^{-i \vect{q}_c \cdot
      \vect{r}_0}\right]\; ,
\end{equation}
and specifically for the two stripe phases:
\begin{align*}
  E_{cl}^{\varphi=0} &= - \Frac{4 \pi^2 D^2 \rho_0}{\lambda_c}
  e^{-2\pi d/\lambda_c} \cos \left(2 \pi \Frac{\bar{x}}{\lambda_c} -2
  \theta\right) \\
  E_{cl}^{\varphi=\pi/2} &= - \Frac{4 \pi^2 D^2 \rho_0}{\lambda_c}
  e^{-2\pi d/\lambda_c} \cos^2 \theta \cos\left(2 \pi
  \Frac{\bar{y}}{\lambda_c}\right) \; .
\end{align*}
Therefore, we can conclude that, for both stripe configurations, the
distance, $\bar{x}$ or $\bar{y}$, that minimizes the interaction
energy $E_{cl}$ does not depend on the dipole strength $D$ or the
layer separation $d$. Further, for the $\varphi=\pi/2$ stripe phase,
the best configuration is the one where the single dipole in layer $2$
aligns with the maximum density of layer $1$, i.e.,
$\bar{y}_{\text{min}}=0$. By contrast, for the $\varphi=0$ stripe
phase, the optimal configuration is for a phase shift equal to twice
the dipole tilt angle $\theta$, i.e., $2\pi
\bar{x}_{\text{min}}/\lambda_c = 2\theta$.
An analogous calculation was carried out for the $\varphi=\pi/2$
stripe phase in the simplified limit where the density modulations were
approximated as discrete dipolar wires~\cite{block2012}.

We now wish to extend these results to multiple layer $N>2$
configurations.
We have seen that the presence of a second layer stabilises the region
of collapse at large tilt angles $\theta_c <\theta < \pi/2$, replacing
it with a novel stripe phase oriented along
$\varphi=0$~\cite{marchetti_13}. Furthermore, within the X-STLS
approximation, the critical tilt angle for the $\varphi=0$ stripe
phase in bilayers is $\theta_c(N=2) \sim 0.79$, which is lower than
that for collapse in the single layer, $\theta_c(N=1) \sim 0.89$.
It is therefore natural to ask whether the $\varphi=0$ stripe phase
will tend to dominate the phase diagram as the number of layers $N$ is
increased.

%%%%%%%%%%%%%%%%%%%%%%%%%%%%%%%%%%%%%%%%%%%%%%%%%%%
\section{$N$ layers}
\label{sec:finit}
Motivated by the results obtained for the bilayer system, we now apply
the X-STLS approximation scheme to the general case of finite $N>2$
layers and evaluate the occurrence of the $\varphi=0$ stripe phase
when varying the system parameters. In particular, by neglecting all
correlations except for the exchange ones, we assume that all
off-diagonal local field factors are zero, $G_{j\ne l}(\vect{q}) = 0$,
while the intralayer ones $G(\vect{q})$ are evaluated according to
Eq.~\eqref{eq:locaz}. To locate the stripe instabilities, we extract
the smallest eigenvalue of the static density-density response
function matrix, $\tilde{\chi}_{\text{min}}(\ve{q})$, and determine
the critical wavevector $\vect{q}_c = (q_c,\varphi_c)$ at which it
first diverges.
If the instability is for a specific angle $\varphi_c$, then it
signals the formation of a density wave with modulations in that
direction and with a period set by $q_c$. Here, we always find that
$\varphi_c = 0$, as in the bilayer case.

At the stripe transition, the phase shifts $\eta_{jl}$ between the
stripes in different layers $j,l$ are extracted from the eigenvector
associated with the smallest eigenvalue,
$\tilde{\chi}_{\text{min}}(\ve{q}_c)$. We find that the behavior of
the multilayer system is a natural extension of the bilayer case: the
phase shift between nearest neighbour layers $\eta_{j j+1}$ grows
monotonically with the tilt angle $\theta$, although linearly only for
small values of $\theta$. Moreover, the phase shift between more
distant layers is always proportional to $\eta_{j j+1}$.

\begin{figure}
\centering
\includegraphics[width=\linewidth,angle=0]{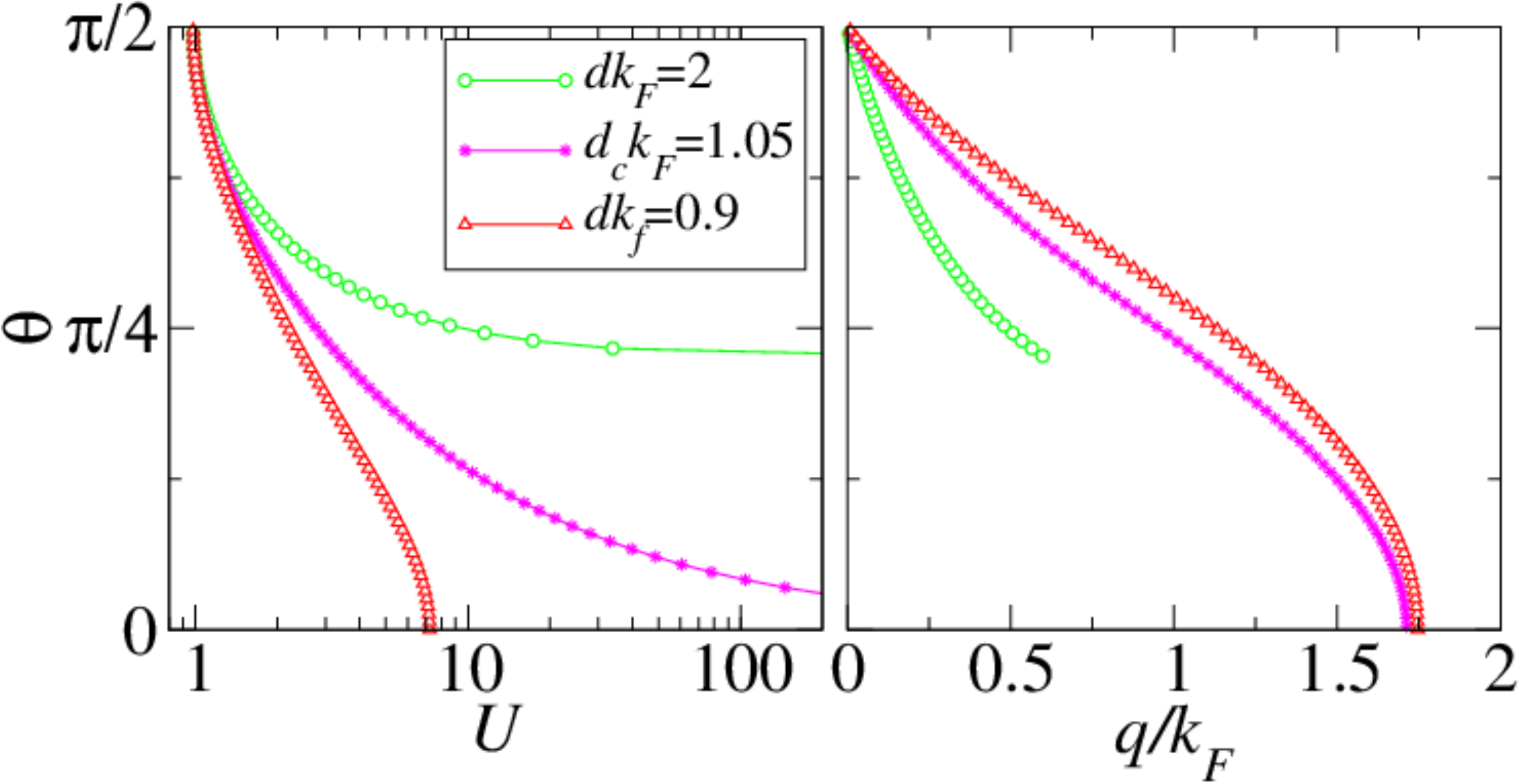}
\caption{(Color online) Evolution of the $\varphi=0$ stripe phase for
  the trilayer $N=3$ geometry when varying the dimensionless layer
  separation $d k_F$. Left panel: phase boundary in the $\theta$
  versus $U$ plane; the instability to the stripe phase is on the
  right side of the plotted boundary. Right panel: Rescaled stripe
  wavevector $q_c/k_F$ at the phase boundaries as a function of the
  tilt angle $\theta$.}
\label{fig:Nequ3}
\end{figure}
\begin{figure}
\centering
\includegraphics[width=\linewidth,angle=0]{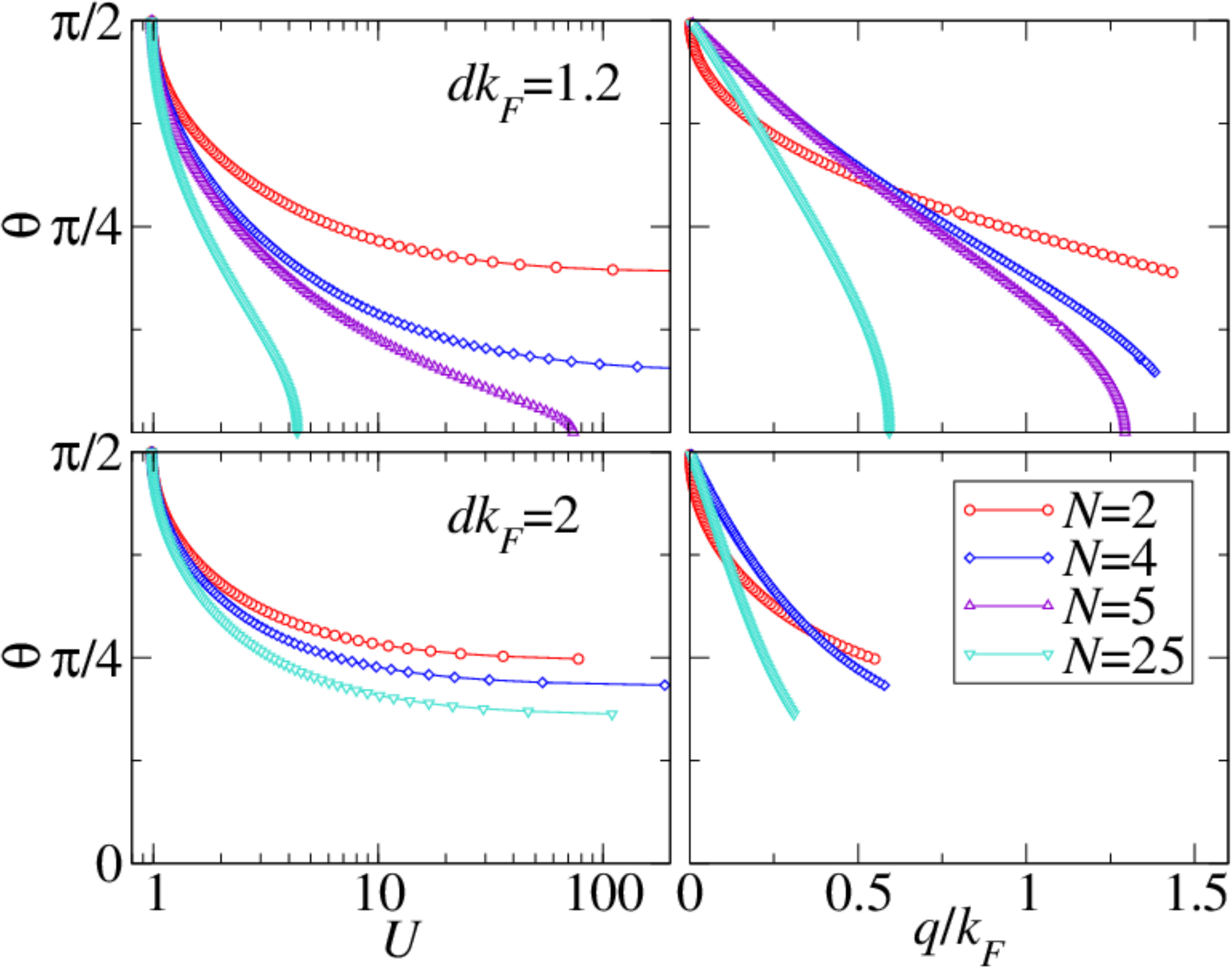}
\caption{(Color online) Phase boundaries (left panels) and rescaled
  stripe wavevector at the boundaries $q_c/k_F$ (right panels) for the
  $\varphi=0$ stripe phase for different values of the number of
  layers $N$ and two fixed values of the layer distance: $d k_F = 1.2$
  (top panels) and $d k_F = 2$ (bottom panels).}
\label{fig:chanN}
\end{figure}
\begin{figure}
\centering
\includegraphics[width=0.95\linewidth,angle=0]{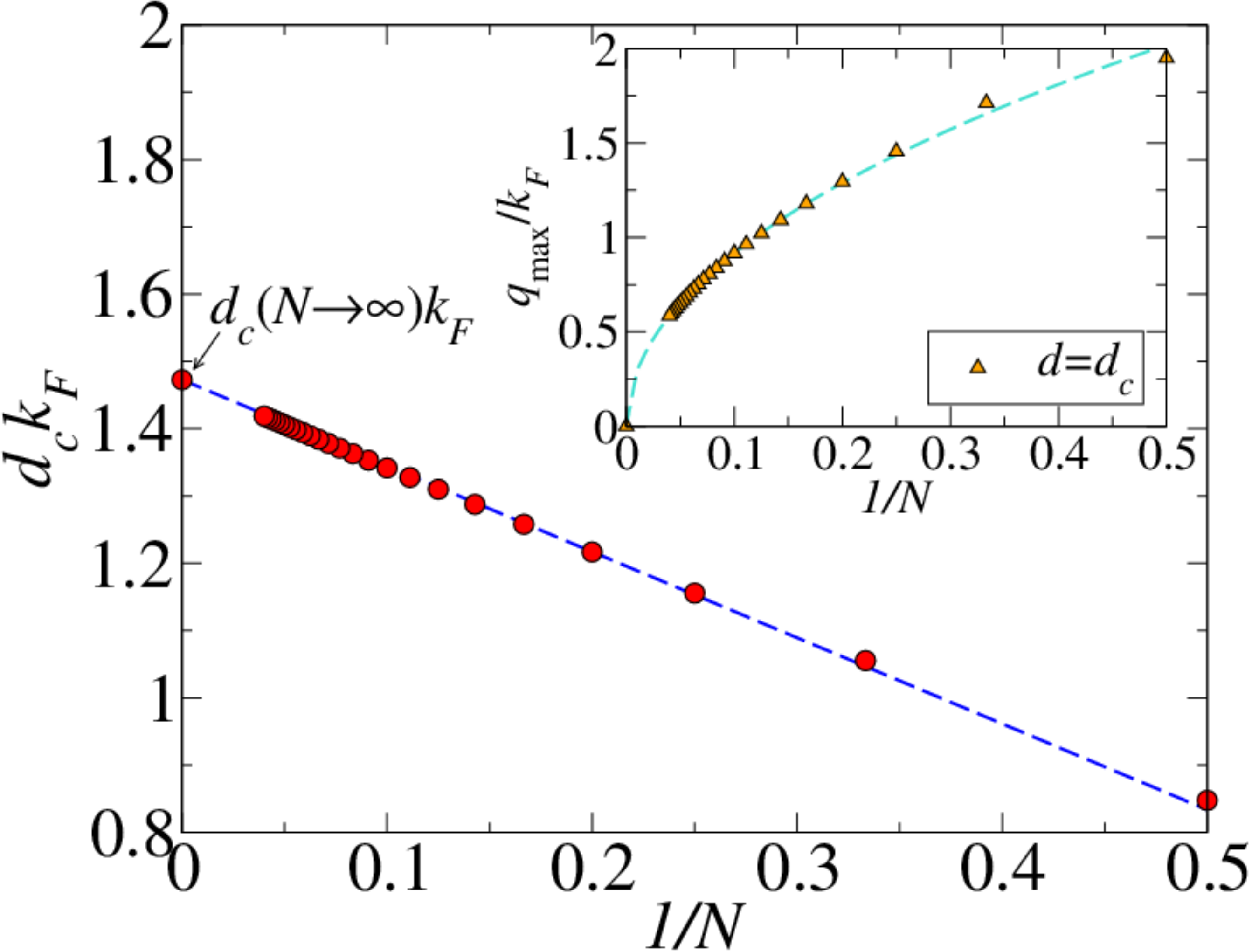}
\caption{(Color online) Critical value of the interlayer distance $d_c
  k_F$ at which $\theta_c=0$ (at $U \to \infty$) as a function of the
  inverse number of layers, $1/N$. In the limit $N \to \infty$, $d_c
  k_F=1.47$. Numerical data are represented as (red) circles, while
  the dashed (blue) line is a linear fit to the data. Inset: maximum
  value of the stripe wavevector $q_c/k_F$ for $d=d_c$ as a function
  of $1/N$. Data are (orange) triangles, while the (turquoise) dashed
  line is a non-linear fit giving $f(1/N)=2.83 (1/N)^{0.49}$.}
\label{fig:dcrit}
\end{figure}

To gain insight into the phase diagram of the multilayer system, we
first focus on the trilayer $N=3$. In Fig.~\ref{fig:Nequ3}, we plot
the phase boundaries for the instability to a $\varphi=0$ stripe phase
(left panel) and the associated critical wavevectors $q_c/k_F$ (right
panel) for different values of the layer distance $dk_F$.
The qualitative behaviour we observe here is common to any number of
layers $N$, including the case of a bilayer $N=2$. For large enough
layer distance $d$, the $\varphi=0$ stripe phase can only occur for
values of the tilt angle greater than a critical angle $\theta_c$,
i.e., at the stripe instability boundary, we have $U \to \infty$ for
$\theta \to \theta_c$ ([green] circle symbols).  However, when the
layer distance decreases, we find that $\theta_c$ eventually reaches
zero at a critical distance $d_c$ ([violet] star symbols): Here, the
$\varphi=0$ stripe phase spans the entire range of tilt angles
$\theta$.  For smaller distances, $d<d_c$, stripe formation is always
possible for sufficiently large but finite values of the interaction
strength $U$, even for dipoles aligned perpendicular to the planes.

Note that when decreasing the value of $d k_F$, eventually our
exchange-only formalism becomes questionable, since it neglects
interlayer correlations. In fact, for $k_F d \lesssim 1$, interlayer
pairing (e.g., dimers in the two-layer
configuration~\cite{volosniev2011} and bound chains in
multilayers~\cite{Volosniev_2013}) is expected to dominate over stripe
formation, and this is not included in our approximation scheme.

We then investigate whether, by fixing the layer distance to a value
$k_F d > 1$, the $\varphi=0$ stripe phase can dominate the phase
diagram as the number of layers $N$ is increased. To this end, we plot
in Fig.~\ref{fig:chanN} the phase boundaries (left panels) for the
$\varphi=0$ stripe phase for different values of $N$. We observe
qualitatively different behaviour depending on whether the distance
$d$ is above or below $\sim1.47/k_F$, corresponding to the critical
distance $d_c$ for $N\to\infty$, as derived in Sec.~\ref{sec:infin}.
When $d>d_c(N \to \infty)$ (lower panels of Fig.~\ref{fig:chanN}), the
$\varphi=0$ stripe phase exists for an increasingly larger range of
dipole tilt angles as $N$ increases, but the critical angle $\theta_c$
finally saturates to a finite positive value. When instead $d<d_c(N
\to \infty)$ (upper panels of Fig.~\ref{fig:chanN}), the stripe phase
eventually spans the full range of angles, including the situation
where the dipole moments are aligned perpendicular to the planes.

These results are summarized in Fig.~\ref{fig:dcrit}, where we plot,
as a function of $1/N$, the critical value of the interlayer distance
$d_c k_F$ at which the $\varphi=0$ stripe phase first spans the full
range of dipole tilt angles.
The data for $N \to \infty$ in the figures are evaluated following the
procedure explained in the next section.

%%%%%%%%%%%%%%%%%%%%%%%%%%%%%%%%%%%%%%%%%%%%%%%%%%%
\section{The $N\to \infty$ limit}
\label{sec:infin}
We now show how the calculation for the $\varphi=0$ stripe instability
can be extended to the limit of an infinite number of layers. The key
point is that the interlayer interaction potential $v_{j l}
(\vect{q})$ only depends on the layer index difference $|j-l|$, so
that, for a system with periodic boundary conditions and $N\gg 1$, we
can make a transformation from the layer index space $j=1, 2, \dots,
N$ to the reciprocal space $p = 2\pi m/N$, where $m = -N/2, \ldots ,
N/2$~\cite{Wigner-CDW}:
\begin{equation}
  \tilde{u}_{p} = \sum_{j-l = -N}^{N} e^{-i (j-l)p} v_{jl} \; .
\end{equation}
Of course, in the actual system, we do not have periodic boundary
conditions, but this should not change the physics in the limit $N \to
\infty$. Inserting Eq.~\eqref{eq:inter} yields the analytical solution
\begin{equation}
  \tilde{u}_{p} (\ve{q}) = - 2\pi D^2q
  \left[\frac{\zeta(\theta,\varphi)}{e^{ip+qd} -1} +
    \frac{\zeta(\theta,\varphi+\pi)}{e^{-ip+qd} -1}\right]\; ,
\end{equation}
where $\zeta(\theta,\varphi) = \xi(\theta,\varphi) + i \sin 2\theta
\cos \varphi$, and where we have taken the limit $N\to \infty$ after
evaluating the geometric series.

We thus obtain the following expression for the eigenvalues of the
inverse density-density response function:
\begin{equation}
  \tilde{\chi}^{-1}_{p} (\ve{q},\omega) = \frac{1}{\Pi
    (\ve{q},\omega)} - v (\ve{q})[1 - G (\ve{q})] -
  \tilde{u}_{p}(\ve{q})\; .
\end{equation}
To investigate the instabilities, we take the static limit, $\omega =
0$, and determine the values of $p$ and $\ve{q}$ for which
$\tilde{\chi}^{-1}_p$ first hits zero. In practice, this means we must find
the value of $p$ that maximizes $- \tilde{u}_{p}(\ve{q})$ for each
$\ve{q}$. Solving for the stationary points gives us two solutions:
\begin{equation}
  p_{i=1,2} = 2 \arctan\left[\frac{\left(e^{qd} \mp 1 \right)
      \sin\theta\cos\varphi}{\left(e^{qd} \pm 1 \right) \cos\theta}
    \right] +\pi \delta_{i,2} \; ,
\end{equation}
where the argument is positive if we assume $0 \leq \varphi \leq
\pi/2$. The first solution $p_1$ corresponds to the maximum of $-
\tilde{u}_{p}(\ve{q})$, where we have
\begin{equation}
  \tilde{u}_{p_1}(\ve{q}) = -4 \pi D^2q \left(\frac{\cos^2
    \theta}{e^{qd} - 1} - \frac{\cos^2 \varphi \sin^2 \theta}{e^{qd}
    +1}\right)\; .
\end{equation}
Thus we have now considerably simplified the problem, as we only have
to find the zero of the maximum inverse eigenvalue of the static
response, $\tilde{\chi}^{-1}_{p_1} (\ve{q}, 0)$, as a function of
$\vect{q}$.

\begin{figure}
\centering
\includegraphics[width=\linewidth,angle=0]{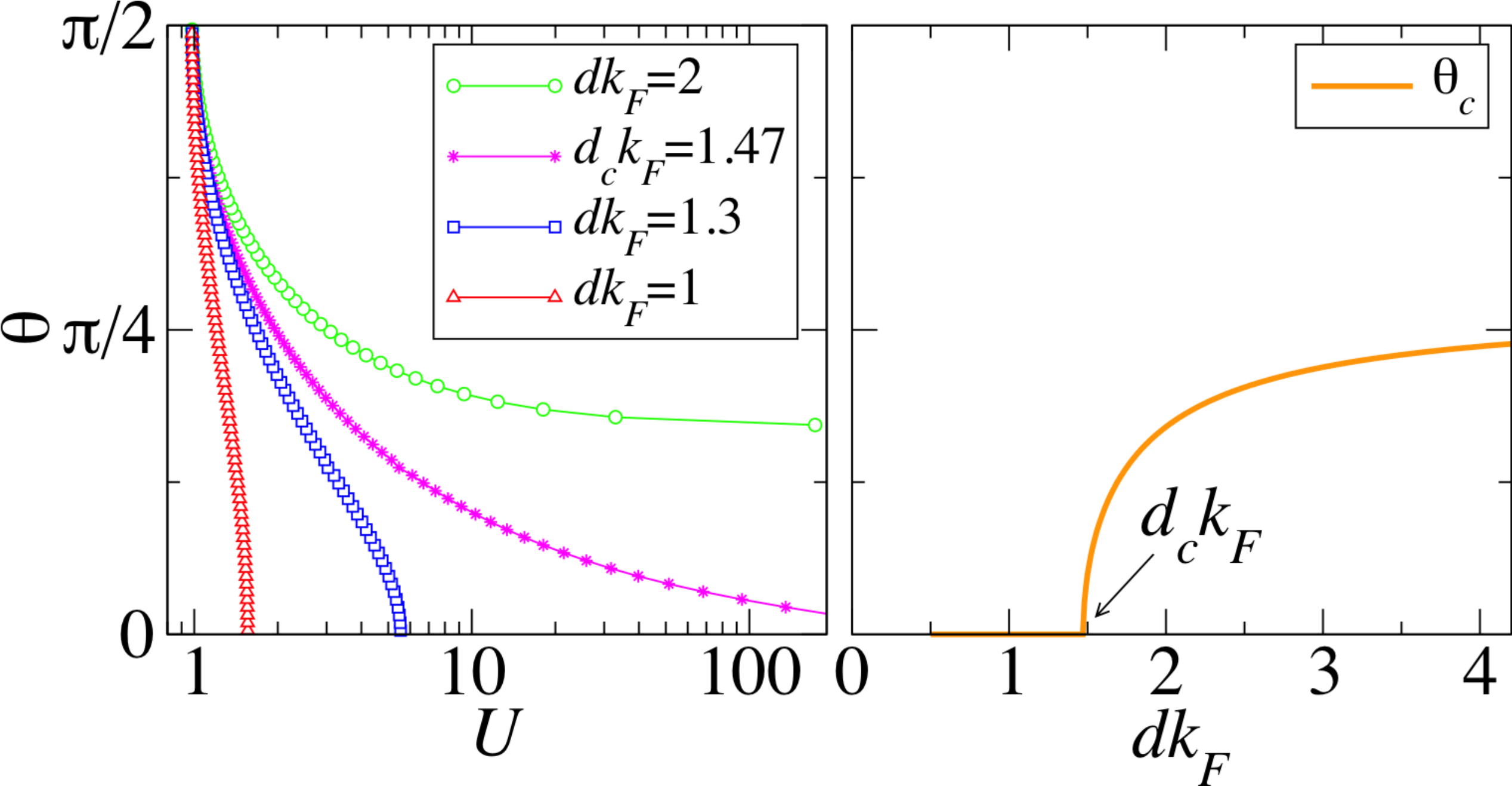}
\caption{(Color online) Left panel: Phase boundary of the collapsed
  region for an infinite number of layers $N \to \infty$ and different
  values of the rescaled layer density $d k_F$. The instability to
  collapse is signalled by an infinite compressibility of the
  dipolar gas, i.e., by a divergence of the static response function
  for $\ve{q}_{c}=0$. While for $d>d_c$, collapse occurs only for a
  tilt angle larger than a critical value $\theta_c$, for $d<d_c$, the
  collapsed phase spans the entire range of angles. Right panel: We
  plot the asymptotic values of the tilt angle at the phase boundary
  for $U \to \infty$, $\theta_c$, as a function of the rescaled layer
  distance $d k_F$; we find that the critical distance for infinite
  layers is given by $d_c (N \to \infty) k_F = 1.47$.  }
\label{fig:Ninfi}
\end{figure}

In this limit, we always find that $(q_c,\varphi_c)=0$, i.e., by
increasing the number of layers to infinity, the gas becomes unstable
towards mechanical collapse. Here, the gas compressibility, which is
proportional to the static response function at $\ve{q}=0$, is
infinite. The results for the phase boundary of such a collapsed phase
are summarized in Fig.~\ref{fig:Ninfi} and are qualitatively similar
to the boundaries found for the $\varphi = 0$ stripe phase in the case
of a finite number of layers. Below the critical layer distance $d_c
k_F \simeq 1.47$, the collapsed phase exists for any value of the
dipole tilt angle, including $\theta=0$.

The behaviour of the infinite $N$ multilayer system is reminiscent of
that expected for the 3D dipolar Fermi gas, where one also has
collapse for sufficiently large dipolar
interactions~\cite{Sogo2009}. In particular, the $\theta = 0$ case
possesses the same rotational symmetry as the 3D gas with aligned
dipole moments. Therefore, it is instructive to compare the onset of
collapse for $\theta=0$ in the multilayer system to that in the 3D
case.  First, one can define a 3D density for the multilayer geometry:
\begin{equation}
  n_{\text{3D}} = \frac{n}{d} = \frac{k_F^3}{4\pi (k_F d)} \; .
\end{equation}
This then yields the corresponding 3D dimensionless interaction
parameter:
\begin{equation}
  U_{\text{3D}} \equiv \frac{mD^2}{\hbar^2} n_{\text{3D}}^{1/3} =
  \frac{U}{(4\pi k_F d)^{1/3}} \; .
\end{equation}
In the 3D dipolar gas, the collapse instability occurs at
$U_{\text{3D}} \simeq 2.4$~\cite{Sogo2009}. Thus, to obtain a
comparable $U_{\text{3D}}$ for collapse in the multilayer system, we
require $d k_F \simeq 1.31$ and $U \simeq 6.12$.
In our 2D infinite layer configuration, we can interpret the parameter
$d k_F$ as an effective Fermi surface ``deformation'' parameter if we
treat $2\pi/d$ as a Fermi momentum in the $z$ direction.  This assumes
that it is the inter-particle spacing rather the fermion exchange that
is the key feature of the Fermi surface deformation in 3D when
considering the collapse instability. For the multilayer geometry, the
ratio between the Fermi momenta in the $z$ and radial directions is
then given by $2 \pi/(k_F d)$, thus yielding $k_{F,z}/k_{F,r} \sim
4.8$ at the collapse instability.  This is not so dissimilar from that
obtained for the 3D Fermi gas within Hartree-Fock mean-field theory,
where $k_{F,z}/k_{F,r} \sim 2$~\cite{Sogo2009}.
In the layered system, we can effectively tune the deformation such
that the critical $U_{\text{3D}}$ for collapse is raised (lowered) by
increasing (decreasing) $dk_F$.  Eventually, when $d>d_c$, the Fermi
surface is not sufficiently elongated along the dipole direction to
produce collapse.

%
%%%%%%%%%%%%%%%%%%%%%%%%%%%%%%%%%%%%%%%%%%%%%%%%%%%
\section{Concluding remarks}
\label{sec:concl}
In this work, we have analysed the density instabilities of dipolar
Fermi multilayer systems that are driven by the attractive part of the
dipolar interaction. We have argued that such instabilities are
dominated by exchange correlations and can thus be described using a
simplified exchange-only STLS approach. We find that the
attraction-driven $\varphi=0$ stripe phase expands to fill the phase
diagram with an increasing number of layers $N$. However, at the same
time, the stripe wavevector decreases so that the stripe phase is
eventually replaced by collapse as $N\to\infty$. For the case
$\theta=0$, the infinite $N$ limit resembles a 3D dipolar gas with a
Fermi surface deformation that can be tuned by varying the interlayer
distance $dk_F$.

Our predicted stripe phases should be assessable in experiments on
polar molecules with sufficiently large dipole moments. One also needs
to consider the issue of losses in experiments when strong
interactions are involved. The restricted motion in the 2D geometry
reduces the possibility of head-to-tail collisions between dipoles,
which underlies the dominant loss process in dipolar gases, but such
collisions are not necessarily suppressed when we have a large dipole
tilt.
However, we expect that chemically stable molecules such as
$^{23}$Na$^{40}$K~\cite{Park_2015} will make it possible to probe this
regime of parameter space.
  
In the future, it would be interesting to extend our results to finite
temperature, where the proliferation of topological defects can melt
the stripe phase~\cite{Wu_2015_b}. Furthermore, one could investigate
the effects of pairing using more sophisticated approaches to the
multilayer system such as the Euler-Lagrange Fermi-hypernetted-chain
approximation~\cite{Abedinpour_2014}.  Finally, there is the
intriguing question of how our predicted phase diagram connects with
other instabilities such as nematic phases or collapse within the
stripe phase.

%%%%%%%%%%%%%%%%%%%%%%%%%%%%%%%%%%%%%%%%%%%%%%%%%
\acknowledgments We are grateful to G. M. Bruun and P. A. Marchetti
for useful discussions.
FMM acknowledges financial support from the Ministerio de Econom\'ia y
Competitividad (MINECO), projects No.~MAT2011-22997 and
No.~MAT2014-53119-C2-1-R.
MMP acknowledges support from the EPSRC under Grant No.\ EP/H00369X/2.
%%%%%%%%%%%%%%%%%%%%%%%%%%%%%%%%%%%%%%%%%%%%%%%%%

%%%%%%%%%%%%%%%%%%%%%%%%%%%%%%%%%%%%%%%%%%%%%%%%%
%%%%%%%%%%%%%%%%%%%%%%%%%%%%%%%%%%%%%%%%%%%%%%%%%
\appendix

\section{Non-interacting static structure factor}
\label{sec:nissf}
\begin{figure}%[ht]
\centering
\includegraphics[width=0.65\linewidth,angle=0]{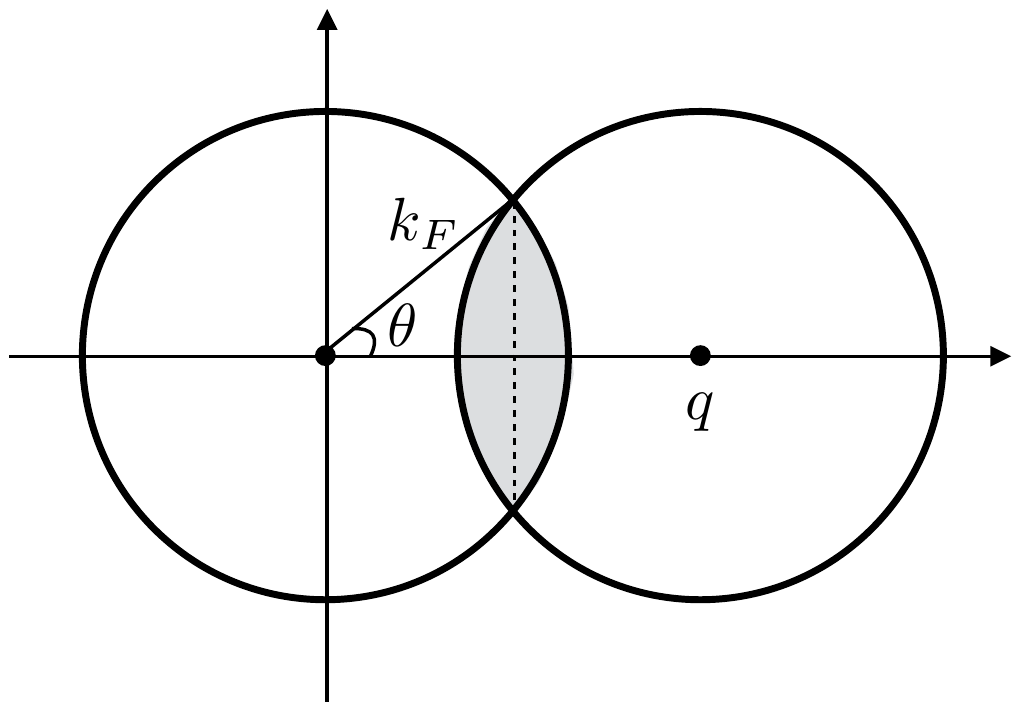}
\caption{Two identical circles of radius $k_F$, where the centers are
  separated by a distance $q$ in $k$-space. The area of the shaded
  overlapping region corresponds to the integral in
  Eq.~\eqref{eq:stati}, defining the the non-interacting static
  structure factor in 2D.}
\label{fig:geome}
\end{figure}

We start with the general expression for the non-interacting static
structure factor in two dimensions~\cite{vignale_book}:
\begin{align}
  S^{(0)}(\ve{q}) = 1 - \frac{1}{n} \int \Frac{d \ve{k}}{(2\pi)^2}
  n_\ve{k} n_{\ve{k}+\ve{q}} \; ,
\label{eq:stati}
\end{align}
where $n$ is the 2D density and $n_\ve{k} = \Theta(k_F - k)$ the
zero-temperature Fermi distribution function. Here, the integral
simply corresponds to calculating the area $A$ of the overlap region
between two identical circles of radius $k_F$, as shown in
Fig.~\ref{fig:geome}. Due to the symmetry of the problem, we only need
to consider half of the overlap region as follows.

Assuming $q < 2 k_F$, we first determine the segment area spanned by
the angle $2\theta$ in the left circle of Fig.~\ref{fig:geome}:
\begin{align*}
  A_{\rm seg} = k_F^2 \theta
\end{align*}
where $\cos\theta = q/2k_F$. Next, we determine the area of the left
triangle obtained by drawing a line through the points where the
circles intersect:
\begin{align*}
  A_{\Delta} = \frac{1}{2} k_F q \sin\theta = \frac{1}{2} k_F q
  \sqrt{1-\left(\frac{q}{2k_F}\right)^2}
\end{align*}
Then we obtain:
\begin{multline} 
  A = 2(A_{\rm seg} - A_{\Delta}) \\
  = 2k_F^2 \arccos\left(\frac{q}{2k_F}\right) - k_F q
  \sqrt{1-\left(\frac{q}{2k_F}\right)^2}
\label{eq:area}
\end{multline}
Inserting~\eqref{eq:area} into~\eqref{eq:stati}, and using the fact
that $\arcsin x = \pi/2 - \arccos x$, we finally recover
Eq.~\eqref{eq:struz} in the main text.

%
%%%%%%%%%%%%%%%%%%%%%%%%%%%%%%%%%%%%%%%%%%%%%%%%%

%\bibliography{dipoleRefs}


\end{document}